%
\magnification = \magstep1
\hsize = 16 truecm 
\parskip = 0pt
\normalbaselineskip = 18pt plus 0.2pt minus 0.1pt
\baselineskip = \normalbaselineskip
%
%

\font\brm=cmr10 scaled \magstep1
\font\bbf=cmbx10 scaled \magstep1
\font\bit=cmti10 scaled \magstep1



%

\def\Bigbreak{\par \ifdim\lastskip < \bigskipamount \removelastskip \fi
                   \penalty-300 \vskip 10mm plus 5mm minus 2mm}
%
%
\newcount\eqnum
\newcount\tempeq
\def\cleareqnum{\global\eqnum = 0}
\def\eqname{(\the\chnum.\the\eqnum)}
\def\neweq{\global\advance\eqnum by 1 \eqno\eqname}
\def\neweqalign{\global\advance\eqnum by 1 &\eqname}
\def\releq#1{\global\tempeq=\eqnum \advance\tempeq by #1
             (\the\chnum.\the\tempeq)}

\cleareqnum
%
%
\newcount\chnum
\def\clearchnum{\global\chnum = 0}
%
%

\def\newchapt#1{\Bigbreak \global\advance\chnum by 1
                \cleareqnum
                \centerline{\bbf\the\chnum.{ }#1}
                \nobreak\vskip 5mm plus 2mm minus 1mm}
\clearchnum
%
%
\newcount\notenumber
\def\clearnotenumber{\notenumber=0} 
\def\note{\advance\notenumber by1 \footnote{$^{\the\notenumber}$}} 
\clearnotenumber 
%
%
\newbox\Eqa
\newbox\Eqb
\newbox\Eqc
\newbox\Eqd
\newbox\Eqe
\newbox\Eqf
\newbox\Eqg
\newbox\Eqh
\newbox\Eqi
\def\storeeq#1{\setbox #1=\hbox{\eqname}}

%
%
\pageno=0
\footline={\ifnum\pageno=0\strut\hfil\else\hfil\tenrm\folio\hfil\strut\fi}%
\def\e{{\rm e}}
\def\H{{\cal H}}
\def\D{{\cal D}}
\def\L{{\cal L}}

\def\E{{\cal E}}
\def\F{{\cal F}} 

\def\P{{\cal P}} 
\def\V{{\cal V}}
\def\W{{\cal W}}

\def\M{{\cal M}}

\def\B{{\cal B}}

\def\R{{\bf R}}
\def\C{{\bf C}}

\def\der{\partial }

\def\a{a(f)}
\def\ac{a^\ast (f)}

\def\xtilde{\widetilde x}

\def\det{ {\rm det }}
\def\alg{{\cal A}_R} 
\def\balg{{\cal B}_R}
\def\balgl{{\cal B}_{R,\lambda }}

\def\brep{{\cal F}_{R,B;m}}

\def\form{\langle \, \cdot \, , \, \cdot \, \rangle_m } 
\def\hl{{\bf R}_+}
%
%

\rightline {hep-th/9607085} 
\centerline {\bbf Boundary Exchange Algebras and} 
\centerline {\bbf Scattering on the Half Line}
\vskip 1 truecm
\centerline {\brm Antonio Liguori} 
\medskip
\centerline {\it Dipartimento di Fisica dell'Universit\`a di Pisa,}
\centerline {\it Piazza Torricelli 2, 56100 Pisa, Italy}
\centerline {\it Istituto Nazionale di Fisica Nucleare, Sezione di Pisa} 
\bigskip 
\centerline {\brm Mihail Mintchev} 
\medskip
\centerline {\it Istituto Nazionale di Fisica Nucleare, Sezione di Pisa}
\centerline {\it Dipartimento di Fisica dell'Universit\`a di Pisa,}
\centerline {\it Piazza Torricelli 2, 56100 Pisa, Italy}
\bigskip
\centerline {\brm Liu Zhao} 
\medskip
\centerline {\it Institute of Modern Physics, 
Northwest University, Xian 710069, China}
\bigskip 
\vskip 1 truecm
\centerline {\bit Abstract} 
\medskip

Some algebraic aspects of field quantization in space-time 
with boundaries are discussed. We introduce an 
associative algebra $\balg $, whose exchange properties are 
inferred from the scattering processes in integrable 
models with reflecting boundary conditions on the half line. 
The basic properties of $\balg $ are established and the Fock 
representations associated with certain involutions in $\balg $ 
are derived. We apply these results for the construction of 
quantum fields and for the study of scattering on the half line. 

\bigskip 
\bigskip  
\vfill \eject


\newchapt{Introduction}

It is well known that the presence of boundaries in space 
affects the behavior of quantum fields. In this paper 
we discuss the influence of the boundary conditions on the 
canonical commutation relations between creation and annihilation 
operators. Our investigation is inspired mainly by the factorized 
scattering theory of integrable models with reflecting boundary 
conditions on the half line. In the absence of boundaries\break 
[6,13,26], the algebraic features of these models are encoded 
in the Zamolodchikov-Faddeev (Z-F) algebra [6,26], denoted in 
what follows by $\alg $. This is an associative algebra, whose 
generators satisfy quadratic constraints, known as exchange 
relations. The Fock representation of $\alg $ equipped 
with an appropriate involution describes the scattering processes 
in integrable models. In this respect one should recall first 
that the Fock space contains two dense subspaces whose elements 
are interpreted as asymptotic in- and out-states. Second, 
the $S$-matrix can be explicitly constructed as a unitary 
operator interpolating between the asymptotic in- and out-spaces. 

In a pioneering paper from the middle of the eighties, 
Cherednik [4] suggested a possible generalization of factorized 
scattering theory to integrable models with reflecting 
boundary conditions, which preserve integrability. The recent 
efforts to gain a deeper insight in various 
boundary-related two-dimensional phenomena, 
stimulated further investigations [5,7-12,16,21,23-25] in this subject. 
Among others, we would like to mention the attempts to 
develop an algebraic approach. One of the basic ideas there 
is to extend the Z-F algebra by introducing [8-12] ``boundary 
creating" (also called ``reflection") operators, which formally 
translate in algebraic terms the nontrivial boundary conditions. 
When possible, such an algebraic formulation is quite attractive 
because the treatment of the boundary conditions in their 
standard analytic form is as a rule a complicated matter. 
In spite of the great progress in implementing the above idea in 
particular models, the fundamental features of the boundary 
operators and their interplay with the ``bulk" theory are still 
to be investigated. This is among the main purposes of the present 
paper. 

We start our analysis by introducing an exchange 
algebra $\balg $ with the following structure. 
In the above spirit, $\balg $ contains both 
boundary and bulk generators. The latter have a counterpart 
in $\alg $, but we shall see that the 
exchange of two bulk generators of $\balg $ 
involves in general boundary elements. The impact of the 
boundary on the bulk theory is therefore manifest 
already on the algebraic level, while the detailed boundary 
conditions are specified on the level of representation. 
We concentrate in this article on the Fock representations of 
$\balg $. We will show that there exist two series of such 
representations, depending on certain involutions in $\balg $. 
We shall construct these representations explicitly, establishing 
also their basic properties. As an application of these results, 
we will perform a detailed and rigorous investigation 
of the $S$-matrix of integrable models in the presence of 
reflecting boundaries. 

The paper is organized as follows. In Sect. 2 we define the 
exchange algebra $\balg $ and investigate some of its basic features. 
We introduce the concept of reflection\break $\balg$-algebra and the 
related notion of reflection automorphism. At the end of this section 
we describe also a family of natural generalizations of $\balg $. 
Sect. 3 is devoted to the Fock representations of $\balg $. In 
Sect. 4 we describe some applications. We show that the second 
quantization on the half line naturally leads to $\balg $. We 
also analyze here the scattering operator of integrable models. 
The last section contains our conclusions. In the appendix we 
construct representations of $\balg $ carrying a boundary quantum 
number.  

This article brings together and extends the results independently 
obtained by the present authors in [19] and [27].

\newchapt {The Exchange Algebra $\balg $}

$\balg $ is by definition an associative 
algebra with identity element $\bf 1$. It has two types of 
generators: 
$$
\{a_\alpha (x),\, a^{\ast \alpha } (x) \, \,  : \, \, 
\alpha = 1,...,N,  \, \, x \in \R^s \} \neweq 
$$
and 
$$
\{b_\alpha^\beta (x) \, \,  : \, \, \alpha , \beta = 1,...,N,  
\, \, x \in \R^s \} \quad , \neweq 
$$ 
which, as mentioned in the introduction, are called 
bulk and boundary generators respectively. For 
convenience, we divide the constraints on (2.1,2) 
in three groups: 

(i) bulk exchange relations are quadratic in the bulk generators 
and read
$$
\eqalignno{
& a_{\alpha_1 }(x_1) \, \, a_{\alpha_2 }(x_2) \, \; - \; \, 
  R_{\alpha_2 \alpha_1 }^{\beta_1 \beta_2 }
  (x_2 , x_1)\,\, a_{\beta_2 }(x_2)\, a_{\beta_1 }(x_1) \, = \, 0 \quad,
  \neweqalign \cr
& a^{\ast \alpha_1 } (x_1)\, a^{\ast \alpha_2 } (x_2) -  
  a^{\ast \beta_2 } (x_2)\, a^{\ast \beta_1 } (x_1)\, 
  R_{\beta_2 \beta_1 }^{\alpha_1 \alpha_2 }(x_2 , x_1) = 0 
  \quad , \neweqalign \cr
& a_{\alpha_1 }(x_1)\, a^{\ast \alpha_2 } (x_2) \; - \; 
  a^{\ast \beta_2 }(x_2)\, 
  R_{\alpha_1 \beta_2 }^{\alpha_2 \beta_1 }(x_1 , x_2)\,  
  a_{\beta_1 }(x_1) = \cr
& \qquad \qquad {1\over 2}\, \delta (x_1 - x_2)\, 
  \delta_{\alpha_1 }^{\alpha_2 }\, {\bf 1} + 
  {1\over 2}\, \delta (x_1 + x_2)\,  b_{\alpha_1 }^{\alpha_2 }(x_1) 
  \quad ; \neweqalign \cr} 
$$

(ii) boundary exchange relations 
$$
R_{\alpha_1 \alpha_2}^{\gamma_2 \gamma_1}(x_1 , x_2)\, 
b_{\gamma_1}^{\delta_1}(x_1)\, 
R_{\gamma_2 \delta_1}^{\beta_1 \delta_2}(x_2 , -x_1)\, 
b_{\delta_2}^{\beta_2}(x_2) = 
$$
$$
b_{\alpha_2}^{\gamma_2}(x_2)\, 
R_{\alpha_1 \gamma_2}^{\delta_2 \delta_1}(x_1 , -x_2)\, 
b^{\gamma_1}_{\delta_1}(x_1)\,
R_{\delta_2 \gamma_1}^{\beta_1 \beta_2}(-x_2 , -x_1) 
\quad ; \neweq 
$$

(iii) mixed relations
$$
a_{\alpha_1}(x_1)\, b_{\alpha_2}^{\beta_2}(x_2) = 
R_{\alpha_2 \alpha_1}^{\gamma_1 \gamma_2}(x_2 , x_1)\, 
b_{\gamma_2}^{\delta_2}(x_2)\, 
R_{\gamma_1 \delta_2}^{\beta_2 \delta_1}(x_1 , -x_2)\, a_{\delta_1}(x_1) 
\quad , \neweq 
$$
$$
b_{\alpha_1}^{\beta_1}(x_1)\, a^{\ast \alpha_2}(x_2) = 
a^{\ast \delta_2}(x_2)\, 
R_{\alpha_1 \delta_2}^{\gamma_2 \delta_1}(x_1 , x_2)\, 
b_{\delta_1}^{\gamma_1}(x_1)\, 
R_{\gamma_2 \gamma_1}^{\beta_1 \alpha_2}(x_2 , -x_1)\, 
\quad . \neweq 
$$

In the above equations and in what follows the summation over 
repeated upper and lower indices is always 
understood. The entries of the exchange factor $R$ are complex 
valued measurable functions on $\R^s \times \R^s$, obeying 
$$
R_{\alpha_1 \alpha_2 }^{\gamma_1 \gamma_2 }(x_1,x_2)\, 
R^{\beta_1 \beta_2 }_{\gamma_1 \gamma_2 }(x_2,x_1) = 
\delta_{\alpha_1 }^{\beta_1 } \, \delta_{\alpha_2 }^{\beta_2 }\quad , \neweq 
$$
$$
R_{\alpha_1 \alpha_2 }^{\gamma_1 \gamma_2 }(x_1,x_2)\,  
R_{\gamma_2 \alpha_3 }^{\delta_2 \beta_3 }(x_1,x_3)\,   
R_{\gamma_1 \delta_2 }^{\beta_1 \beta_2 }(x_2,x_3) = 
R_{\alpha_2 \alpha_3 }^{\gamma_2 \gamma_3 }(x_2,x_3)\,  
R_{\alpha_1 \gamma_2 }^{\beta_1 \delta_2 }(x_1,x_3)\,  
R_{\delta_2 \gamma_3 }^{\beta_2 \beta_3 }(x_1,x_2)  \, .
\neweq
$$
These compatibility conditions are assumed throughout 
the paper and can be considered as general requirements on $R$, 
which together with eqs.(2.3-8) define the exchange algebra $\balg $. 
Eq.(2.10) is the spectral quantum Yang-Baxter equation in 
its braid form, $\R^s$ playing the role of spectral set. 

Let us comment now on the exchange relations (2.3-8), which may look 
at first sight a bit complicated. Concerning the general structure, 
we observe that after setting formally all boundary generators in 
(2.3-8) to zero and rescaling by a factor of 
$1/{\sqrt 2}$ the bulk generators, one 
gets the Z-F algebra $\alg $. This fact clarifies 
partially the origin of eqs.(2.3-5). The presence of boundary 
generators in the right hand side of (2.5) is worth stressing. 
This is one of the essential points, in which our approach 
differs from the previous attempts to define a boundary 
exchange algebra. 

Eq.(2.6) describes the exchange of two boundary generators 
taken in generic points and also deserves a remark. 
It looks similar to the boundary Yang-Baxter 
equation [4]; the difference is that the elements 
$\{b_\alpha^\beta (x) \}$ do not commute 
in general and consequently their position 
in (2.6) is essential. Notice also that $\{b_\alpha^\beta (x) \}$ 
close a subalgebra of $\balg $, which presents by itself some 
interest [24]. Finally, eqs.(2.7,8) express 
the interplay between $\{a_\alpha (x),\, a^{\ast \alpha }(x) \}$ and 
$\{b_\alpha^\beta (x) \}$ and represent another relevant new 
aspect of our proposal.  
 
Two straightforward examples, denoted by $\B_\pm $, correspond 
to the constant solutions 
$$
R_{\alpha_1 \alpha_2 }^{\beta_1 \beta_2 } = 
\pm \, \delta_{\alpha_1}^{\beta_2}\, \delta_{\alpha_2}^{\beta_1}  
\neweq 
$$
of (2.9,10) and represent in the above context the counterparts 
of the canonical\break (anti)commutation relations. Eqs.(2.6-8) 
imply that $\{b_\alpha^\beta (x) \}$ are central elements in 
$\B_\pm $. Nevertheless, also in these relatively simple cases 
the right hand side of eq.(2.5) keeps trace of the nontrivial 
boundary conditions. Two applications of $\B_+$ with $N=1$ 
are described in Sect. 4. 

To further understand the structure of $\balg $ and its representations, 
it is instructive to introduce some involutions in $\balg $. Let 
$H_N$ be the family of invertible Hermitian 
$N\times N$ matrices and let $\M$ be the set of matrix valued functions 
$m\, :\, \R^s \rightarrow H_N$, such that the entries of 
$m(x)$ and $m(x)^{-1}$ are measurable and bounded in $\R^s $. 
Consider the mapping $I_m$ defined by 
$$
I_m \, : \,  a^{\ast \alpha }(x) \longmapsto 
m^\beta_\alpha (x)\, a_\beta (x) \quad , \neweq 
$$
$$
I_m \, : \,  a_\alpha (x) \longmapsto 
a^{\ast \beta }(x)\, 
m^{-1 \alpha}_{\, \, \, \, \, \, \, \beta}(x)  
\quad , \neweq 
$$
$$
I_m \, : \,  b_\alpha^\beta  (x) \longmapsto 
m_\beta^\gamma (-x)\, b_\gamma^\delta (-x)\, 
m^{-1 \alpha}_{\, \, \, \, \, \, \, \delta}(x)  
\quad . \neweq 
$$
Provided that $m\in \M$ satisfies 
$$
R^{\dagger \gamma_1 \gamma_2}_{\, \, \alpha_1 \alpha_2}(x_1 , x_2 )\, 
m^{\beta_1}_{\gamma_1}(x_1 )\, m^{\beta_2}_{\gamma_2}(x_2 ) 
= m^{\gamma_1}_{\alpha_1}(x_2 )\, m^{\gamma_2}_{\alpha_2}(x_1 )\, 
R^{\beta_1 \beta_2}_{\gamma_1 \gamma_2}(x_2 , x_1 ) 
\quad , \neweq 
$$
it is not difficult to check that when extended as an antilinear 
antihomomorphism on $\balg $, $I_m$ defines an involution. In eq.(2.15) 
and in what follows the dagger stands for Hermitian conjugation, i.e.  
$$
R^{\dagger \beta_1 \beta_2 }_{\, \, \alpha_1 \alpha_2} (x_1,x_2) 
\equiv {\overline R_{\beta_1 \beta_2 }^{\alpha_1 \alpha_2}} (x_1,x_2) 
\quad , 
$$
the bar indicating complex conjugation. Notice that for the 
algebras $\B_\pm $ eq.(2.15) is satisfied for any $m\in \M$. 

In this paper we shall concentrate on the following specific type of 
$\balg $-algebras. We call the boundary generators $\{b_\alpha^\beta (x) \}$ 
reflections if  
$$
b_\alpha^\gamma (x)\, b_\gamma^\beta (-x) = \delta_\alpha^\beta  \neweq 
$$
hold. In this case we refer to $\balg $ as reflection exchange algebra. 
The condition (2.16) is $I_m$-invariant and one easily proves   
\medskip 
\noindent {\bf Proposition 1.} {\it Let} $\balg $ {\it be a reflection 
exchange algebra. Then the mapping} 
$$
\varrho \, :\, a_\alpha (x) \longmapsto b_\alpha^\beta (x) a_\beta (-x ) 
\quad , \neweq 
$$
$$
\varrho \, :\, a^{\ast \alpha }(x) \longmapsto a^{\ast \beta } (-x ) 
b^\alpha_\beta (-x) 
\quad , \neweq 
$$
$$
\varrho \, :\, b_\alpha^\beta (x) \longmapsto b^\beta_\alpha (x) 
\quad , \neweq 
$$ 
{\it leaves invariant the constraints} (2.3-8) {\it and extends therefore 
to an automorphism on} $\balg $. {\it Moreover, being compatible with} 
$I_m$, $\varrho $ {\it is actually an automorphism of} $\{\balg , I_m\}$ 
{\it considered as an algebra with involution.} 
\medskip 

In what follows $\varrho $ is called the reflection automorphism 
of $\balg $. Besides encoding some essential features of any reflection 
exchange algebra, $\varrho$ has a direct physical interpretation 
in scattering theory: it provides a mathematical description of 
the intuitive physical picture that bouncing back from a wall, 
particles change the sign of their rapidities. In fact, 
the two elements $a^{\ast \alpha }(-x)$ and 
$a^{\ast \beta } (x) b^\alpha_\beta (x) $ are $\varrho $-equivalent, 
$$
a^{\ast \alpha }(-x) \sim a^{\ast \beta } (x) b^\alpha_\beta (x) 
\quad . \neweq 
$$
This relation in our framework is the counterpart of 
a heuristic equation (see for example eq.(3.22) of [10]), 
conjectured in all papers dealing with factorized 
scattering with reflecting boundaries. In 
the next section we will show that the $\varrho $-equivalence 
becomes actually an equality in the Fock representation of 
$\{\balg , I_m\}$. For proving this statement we will use 
the relations
$$
\left \{a_\alpha (x_1) - \varrho \left [a_\alpha (x_1)\right ]\right \} 
a^{\ast \beta }(x_2) = a^{\ast \gamma }(x_2)\,  
R_{\alpha \gamma }^{\beta \delta }(x_1 ,x_2) 
\left \{a_\delta (x_1) - \varrho \left [a_\delta (x_1)\right ]\right \} 
\quad , \neweq 
$$
$$
\left \{a^{\ast \alpha }(x_1) -
\varrho \left [a^{\ast \alpha }(x_1)\right ] \right \}
a^{\ast \beta }(x_2) 
= a^{\ast \gamma }(x_2) \left \{a^{\ast \delta }(x_1) - 
\varrho \left [a^{\ast \delta }(x_1)\right ] \right \} 
\, R^{\alpha \beta }_{\gamma \delta }(x_2 ,x_1)
\quad , \neweq 
$$ 
whose validity follows directly from eqs.(2.3-5,7,8,16). 

Before concluding this section, we would like to introduce a whole 
class of more general exchange algebras which can be treated in 
the above way. The idea is to replace the reflection 
$x\mapsto -x$, which plays a special role in defining $\balg $, 
with any almost everywhere differentiable 
mapping $\lambda \, :\, x\mapsto {\xtilde} $, satisfying 
the iterative functional equation 
$$
\lambda \left ( \lambda (x) \right ) = x \quad . \neweq 
$$
The resulting exchange algebras will be denoted by $\balgl $ and are 
characterized by the following constraints: the relations (2.3,4) remain 
unchanged, whereas (2.5-8) take the form 
$$
\eqalign{
& a_{\alpha_1 }(x_1)\, a^{\ast \alpha_2 } (x_2) \; - \; 
  a^{\ast \beta_2 }(x_2)\, 
  R_{\alpha_1 \beta_2 }^{\alpha_2 \beta_1 }(x_1 , x_2)\,  
  a_{\beta_1 }(x_1) = \cr
& {1\over 2 |\det \lambda'(x_1)|^{1/2}}\, 
\bigl \{  \delta (x_1 - x_2)\, 
  \delta_{\alpha_1 }^{\alpha_2 }\, {\bf 1} + 
  \delta (x_1 - \xtilde_2)\,  b_{\alpha_1 }^{\alpha_2 }(x_1) \bigr \}
  \quad , \cr} 
\neweq
$$
\smallskip 
$$
\eqalign{ 
&R_{\alpha_1 \alpha_2}^{\gamma_2 \gamma_1}(x_1 , x_2)\, 
b_{\gamma_1}^{\delta_1}(x_1)\, 
R_{\gamma_2 \delta_1}^{\beta_1 \delta_2}(x_2 , \xtilde_1)\, 
b_{\delta_2}^{\beta_2}(x_2) = \cr 
&b_{\alpha_2}^{\gamma_2}(x_2)\, 
R_{\alpha_1 \gamma_2}^{\delta_2 \delta_1}(x_1 , \xtilde_2)\, 
b^{\gamma_1}_{\delta_1}(x_1)\, 
R_{\delta_2 \gamma_1}^{\beta_1 \beta_2}(\xtilde_2 , \xtilde_1)  
\quad , \cr}
\neweq 
$$
\smallskip
$$
\eqalign{ 
&a_{\alpha_1}(x_1)\, b_{\alpha_2}^{\beta_2}(x_2) = 
R_{\alpha_2 \alpha_1}^{\gamma_1 \gamma_2}(x_2 , x_1)\, 
b_{\gamma_2}^{\delta_2}(x_2)\, 
R_{\gamma_1 \delta_2}^{\beta_2 \delta_1}(x_1 , \xtilde_2)\, a_{\delta_1}(x_1) 
\quad , \cr
&b_{\alpha_1}^{\beta_1}(x_1)\, a^{\ast \alpha_2}(x_2) = 
a^{\ast \delta_2}(x_2)\, 
R_{\alpha_1 \delta_2}^{\gamma_2 \delta_1}(x_1 , x_2)\, 
b_{\delta_1}^{\gamma_1}(x_1)\, 
R_{\gamma_2 \gamma_1}^{\beta_1 \alpha_2}(x_2 , \xtilde_1) 
\quad .\cr} 
\neweq 
$$
Here $\lambda'(x)$ denotes the Jacobian matrix of the function 
$\lambda$. The results of this section regarding $\balg $ can 
be transferred with obvious modifications to $\balgl $. For the 
complete set of solutions of eq.(2.23) we refer to [14]. 
When $s=1$ for instance, the mapping $\lambda $ 
can be any almost everywhere differentiable function 
in $\R$ whose graph is symmetric with respect 
to the diagonal $\{(x,y)\in \R^2 \, :\, x=y \}$.  

Summarizing, we introduced so far the exchange algebra 
$\balg $ and some natural generalizations of it. We defined 
also a set of involutions in $\balg $, which are useful in 
representation theory. Focusing on reflection type 
$\balg$-algebras, we shall construct in the next section the 
relative Fock representations.

\newchapt {Fock Representations}

We consider in this paper representations of 
$\{\balg , I_m\}$ with the following general structure. 
\item {1.} The representation space $\L$ is 
a locally convex and complete topological linear space over $\C$. 
\item {2.} The generators 
$\{a_\alpha (x),\, a^{\ast \alpha } (x),\, b_\alpha^\beta (x) \}$ 
are operator valued distributions with common and invariant dense 
domain $\D\subset \L$, where eqs.(2.3-8) hold.  
\item {3.} $\D$ is equipped with a nondegenerate sesquilinear form 
(inner product) $\form $, which is at least separately continuous. 
The involution $I_m$ defined by eqs.(2.12-14) is realized as a 
conjugation with respect to $\form $. 

\noindent A Fock representation of $\{\balg , I_m\}$ is 
specified further by the following requirement. 
\item {4.} There exists a vector (vacuum state) $\Omega \in \D$ 
which is annihilated by $a_\alpha (x)$. Moreover, $\Omega $ is 
cyclic with respect to $\{a^{\ast \alpha } (x)\}$ and 
$\langle \Omega\, ,\, \Omega \rangle_m\, = 1$. 

\noindent A more general situation, when a boundary quantum number 
[10] is present, is outlined in the appendix. 

There is a series of direct but quite important corollaries from the above 
assumptions. Let us start with  
\medskip 
\noindent {\bf Proposition 2.} {\it The automorphism} $\varrho$ 
{\it of any reflection} $\balg$-{\it algebra is implemented in the 
above Fock representations by the identity operator.}
\medskip
\noindent {\it Proof.} First of all we observe that 
$$
\langle P^\prime [a^{\ast}]\Omega \, , \, \{a_\alpha (x) - 
\varrho [ a_\alpha (x)]\} P[a^{\ast}]\Omega \rangle_m = 0
\quad, \neweq
$$
where $P$ and $P^\prime $ are arbitrary polynomials. In fact, by means of 
eq.(2.21) one can shift the curly bracket to the vacuum and use that 
$a_\alpha (x)$ annihilate $\Omega $. Now the cyclicity of $\Omega $, 
combined with the properties of $\form $, allow to 
replace $P^\prime [a^{\ast}]\Omega $ by an arbitrary state 
$\varphi \in \D $. A further conjugation leads to 
$$
\langle P[a^{\ast}] \Omega \, , \, \{a^{\ast \alpha }(x) - 
\varrho [a^{\ast \alpha }(x)] \}\varphi \rangle_m = 0
\quad, \neweq
$$
which implies 
$$
a^{\ast \alpha }(x) = a^{\ast \beta } (-x )b^\alpha_\beta (-x)  \neweq 
$$
on $\D$. Analogously, employing (2.22) one concludes that 
$$
a_\alpha (x) = b_\alpha^\beta (x) a_\beta (-x ) 
\neweq  
$$
also holds on $\D$. Finally, taking in consideration eq.(2.19) 
we deduce that $\varrho $ is indeed implemented by the identity operator. 

{}For describing some further characteristic features of the Fock 
representations of $\balg $, we introduce the c-number distributions
$$
B_\alpha^\beta (x) \equiv \langle 
\Omega \, ,\, b_\alpha^\beta (x)\Omega \rangle_m 
\quad . \neweq 
$$
The requirement 3 implies that 
$$
B_{\, \, \, \alpha}^{\dagger \beta }(x) = m_\alpha^\gamma (-x)\, 
B_\gamma^\delta (-x)\, m^{-1 \beta}_{\, \, \, \, \, \, \, \, \delta}(x) 
\quad , \neweq 
$$
which is the analog of condition (2.15) regarding the 
exchange factor $R$. 

Two other simple consequences of our assumptions 1-4 above are 
collected in 
\medskip
\noindent {\bf Proposition 3.} {\it The vacuum vector} $\Omega $ 
{\it is unique} ({\it up to a phase factor}) {\it and satisfies} 
$$
b_\alpha^\beta (x)\, \Omega =  B_\alpha^\beta (x)\, \Omega 
\quad . \neweq 
$$
\medskip
\noindent {\it Proof.} The argument implying the uniqueness of 
the vacuum is standard. Concerning eq.(3.7), it can be 
inferred from the identity 
$$
\langle [b_\alpha^\beta (x) - B_\alpha^\beta (x)]\Omega \, , 
\, P[a^\ast ]\Omega \rangle_m = 0 
\quad , \neweq 
$$
$P$ being an arbitrary polynomial. In order to prove eq.(3.8) we 
shift by a conjugation the polynomial to the first factor in the 
right hand side of (3.8) and apply afterwards the 
exchange relation (2.8) and eq.(3.5). For 
completing the proof, one also employs that $\Omega $ is 
cyclic and $\form $ is continuous and nondegenerate. 

Combining eq.(3.7) with the fact that $a_\alpha (x)$ 
annihilate $\Omega $, we conclude that\break eqs.(2.5,7,8) 
allow for a purely algebraic derivation of the vacuum expectation 
values involving any number and combination of the generators 
$\{a_\alpha (x),\, a^{\ast \alpha } (x),\, b_\alpha^\beta (x) \}$. 
In particular, taking the vacuum expectation value of eq.(2.6) 
one gets 
$$
\eqalign{
&R_{\alpha_1 \alpha_2}^{\gamma_2 \gamma_1}(x_1 , x_2)\, 
 B_{\gamma_1}^{\delta_1}(x_1)\, 
 R_{\gamma_2 \delta_1}^{\beta_1 \delta_2}(x_2 , -x_1)\, 
 B_{\delta_2}^{\beta_2}(x_2) = \cr
&B_{\alpha_2}^{\gamma_2}(x_2)\, 
 R_{\alpha_1 \gamma_2}^{\delta_2 \delta_1}(x_1 , -x_2)\, 
 B^{\gamma_1}_{\delta_1}(x_1)\,
 R_{\delta_2 \gamma_1}^{\beta_1 \beta_2}(-x_2 , -x_1) 
 \quad .} \neweq 
$$
We thus recover at the level of Fock representation the original 
boundary Yang-Baxter equation [4]. In addition, when one is 
dealing with reflection algebras, eq.(2.17) implies
$$ 
B_\alpha^\gamma (x)\, B_\gamma^\beta (-x) = \delta_\alpha^\beta  
\quad . \neweq 
$$
In this case we refer to $B$ as reflection matrix. 

A final comment in this introductory part concerns the algebras 
$\B_\pm $. Using that $\{b_\alpha^\beta (x)\}$ are 
central elements, in a Fock representation of $\B_\pm$ one has 
$$
b_\alpha^\beta (x)\varphi = B_\alpha^\beta (x)\varphi 
\neweq  
$$
{}for any $\varphi \in \D$. 

At this stage it is convenient to introduce the set $\M(R,B)$ of 
all elements of $\M$ obeying both eqs.(2.15) and (3.6). Then the basic 
input fixing a Fock representation of the reflection algebra $\{\balg , I_m\}$ 
is the triplet $\{R,\, B;\, m\}$, where $R$ and $B$ satisfy eqs.(2.9,10) and 
(3.9,10), and $m\in \M(R,B)$. Some explicit examples of such triplets have 
been found already by Cherednik [4]. With any 
$\{R,\, B;\, m\}$ we associate a Fock representation 
denoted by $\brep $. To the end of this section we will describe 
the explicit construction of $\brep $. 

Our first step will be to introduce the $n$-particle subspace 
$\H_{R,B}^n$ of $\brep $. For this purpose we consider   
$$
\H = \bigoplus_{\alpha =1}^N L^2(\R^s) \quad , \neweq 
$$
equipped with the standard scalar product 
$$
(\varphi , \psi ) = 
\int d^sx \varphi^{\dagger \alpha } (x) \psi_\alpha (x) = 
\sum_{\alpha =1}^N \int d^sx {\overline \varphi }_\alpha (x) \psi_\alpha (x) 
\quad . \neweq 
$$
{}For $n\geq 1$ the $n$-particle space $\H_{R,B}^n$ we are looking for, 
will be a subspace of the $n$-fold tensor power $\H^{\otimes n}$, 
characterized by a suitable projection operator $P_{R,B}^{(n)}$. 
The ingredients for constructing  $P_{R,B}^{(n)}$ 
are essentially two: a specific finite group and its representation 
in $\H^{\otimes n}$, defined in terms of the exchange factor $R$ and 
the reflection matrix $B$. 

Let us concentrate first on the group. In the 
case of $\alg$, this was [17] simply the permutation group $\P_n$. 
The physics behind $\balg $ suggest to enlarge in this case 
the group by adding a reflection generator. More precisely, we 
consider the group $\W_n$ generated by 
$\{\tau ,\, \sigma_i \, \, :\, \, i=1,..., n-1\}$ which satisfy 
$$
\eqalign{
& \sigma_i\, \sigma_j = 
  \sigma_j\, \sigma_i \quad , \quad \quad |i-j|\geq 2 \quad ,   \cr
& \sigma_i\, \tau = 
  \tau \, \sigma_i \quad , \quad \quad 1\leq i<n-2 \quad , \cr}
  \neweq  
$$
$$
\eqalign{
\sigma_i\, \sigma_{i+1}\, \sigma_i  & = \sigma_{i+1}\, \sigma_i\,
\sigma_{i+1} \quad ,  \cr
\sigma_{n-1} \, \tau \, \sigma_{n-1} \, \tau &= 
\tau \, \sigma_{n-1} \, \tau \, \sigma_{n-1} \quad , \cr} 
\neweq 
$$
$$
\sigma_i^2 = \tau^2 = {\bf 1} \quad .\neweq 
$$
$\W_n$ is the Weyl group associated with the root systems of 
the classical Lie algebra $B_n$ and has $2^n n!$ 
elements. Although it contains no permutations, 
$\W_1 = \{{\bf 1}, \tau \}$ is nontrivial. 

We turn now to the representation of $\W_n$ in  
$\H^{\otimes n}$. Observing that any element $\varphi\in \H^{\otimes n}$ 
can be viewed as a column whose entries are 
$\varphi_{\alpha_1 \cdots \alpha_n} (x_1,\dots,x_n)$, we 
define the operators $\{T^{(n)} ,\, S^{(n)}_i \, \, :\, \, i=1,..., n-1\}$ 
acting on $\H^{\otimes n}$ according to: 
$$
\left [S^{(n)}_i\varphi \right ]_{\alpha_1 ... \alpha_n } 
(x_1,...,x_i,x_{i+1},...,x_n ) = 
$$
$$
\left [ R_{i\,i+1}(x_i,x_{i+1})
\right ]_{\alpha_1 ... \alpha_n }^{\beta_1 ... \beta_n } 
\varphi_{\beta_1 ... \beta_n } 
(x_1,...,x_{i+1},x_i,...,x_n ) \quad , \quad \quad n\geq 2 \quad , \neweq 
$$
$$
\left[T^{(n)}\varphi \right ]_{\alpha_1 ... \alpha_n } 
(x_1,...,x_n ) =
$$
$$ 
\left [ B_n(x_n)
\right ]_{\alpha_1 ... \alpha_n }^{\beta_1 ... \beta_n } 
\varphi_{\beta_1 ... \beta_n } 
(x_1,...,x_{n-1},-x_n ) \quad , \quad \quad n\geq 1 \neweq 
$$
where  
$$
\left [ R_{ij}(x_i,x_j)
\right ]_{\alpha_1 ... \alpha_n }^{\beta_1 ... \beta_n } = 
                 \delta_{\alpha_1}^{\beta_1} \delta_{\alpha_2}^{\beta_2} 
                  \cdots { \widehat {\delta_{\alpha_i}^{\beta_i}}}
                  \cdots { \widehat {\delta_{\alpha_j}^{\beta_j}}}
                  \cdots \delta_{\alpha_n}^{\beta_n}
   \,  R_{\alpha_i \alpha_j}^{\beta_i \beta_j}(x_i,x_j) \neweq
$$
and   
$$
\left [B_i(x)\right ]_{\alpha_1 ... \alpha_n }^{\beta_1 ... \beta_n } = 
                 \delta_{\alpha_1}^{\beta_1} \delta_{\alpha_2}^{\beta_2} 
                  \cdots { \widehat {\delta_{\alpha_i}^{\beta_i}}}
                  \cdots \delta_{\alpha_n}^{\beta_n}
   \,  B_{\alpha_i}^{\beta_i}(x) \quad . \neweq
$$
The hat in eqs.(3.19,20) indicates that the corresponding symbol must 
be omitted. For implementing eqs.(3.17,18) on the whole $\H^{\otimes n}$, 
we assume at this stage that the matrix elements 
$R_{\alpha_1 \alpha_2 }^{\beta_1 \beta_2 }(x_1,x_2)$ and 
$B_\alpha^\beta (x)$ are bounded functions. We are now in position 
to prove 
\medskip
\noindent {\bf Proposition 4}: 
$\{T^{(n)} , \,  S^{(n)}_i \, \, :\, \, i=1,...,n-1\, \}$ 
{\it are bounded operators on $\H^{\otimes n}$ and the mapping} 
$$
\chi^{(n)}\, :\, \tau \longmapsto T^{(n)} \quad , 
\quad \quad 
\chi^{(n)}\, :\, \sigma_i \longmapsto S^{(n)}_i \quad , 
\quad \quad i=1,\cdots,n-1 \neweq
$$
{\it defines a representation of} $\W_n$ {\it in} $\H^{\otimes n}$. 
{\it Moreover,} 
$$
P_{R,B}^{(n)} \equiv {1\over 2^n n!}\, 
\sum_{\nu \in \W_n} \, \chi^{(n)}(\nu )  \neweq 
$$
{\it is a bounded projection operator in} $\H^{\otimes n}$. 
\medskip 
\noindent {\it Proof.} The main point is to show that 
$\{T^{(n)},\, S^{(n)}_i \, \, :\, \, i=1,...,n-1\, \}$ obey eqs.(3.14-16). 
This can be checked directly. Eqs.(3.14) are satisfied by construction. 
Eqs.(3.15) follow from (2.10) and (3.9). Finally, eqs.(2.9) and (3.10) 
imply (3.16). 
\medskip 

Let us observe in passing that $P_{R,B}^{(n)}$ is an orthogonal 
projector only if the $N\times N$ identity matrix $e$ belongs 
to $\M(R,B)$. In general $P_{R,B}^{(n)}$ is not orthogonal, 
but being a bounded operator determines for any $n\geq 1$ a 
(nonempty) closed subspace 
$$
\H_{R,B}^n \equiv P_{R,B}^{(n)}\, \H^{\otimes n} \quad . \neweq 
$$
By construction the elements of $\H_{R,B}^n$ behave as follows:   
$$
\varphi_{\alpha_1 ... \alpha_n }(x_1,...,x_i,x_{i+1},...,x_n ) = 
\left [ R_{i\,i+1}(x_i,x_{i+1})
\right ]_{\alpha_1 ... \alpha_n }^{\beta_1 ... \beta_n } 
\varphi_{\beta_1 ... \beta_n } 
(x_1,...,x_{i+1},x_i,...,x_n ) \, \,  ,\neweq 
$$
$$
\varphi_{\alpha_1 ... \alpha_n } (x_1,...,x_n ) = 
\left [ B_n(x_n)
\right ]_{\alpha_1 ... \alpha_n }^{\beta_1 ... \beta_n } 
\varphi_{\beta_1 ... \beta_n } 
(x_1,...,x_{n-1},-x_n ) \quad . \neweq 
$$

Setting $\H_{R,B}^0 = \C^1$, we introduce also the finite 
particle space $\F_{R,B;m}^0(\H )$ 
as the (complex) linear space of sequences 
$\varphi = \left ( \varphi^{(0)}, \varphi^{(1)},...,\varphi^{(n)},...\right )$ 
with $\varphi^{(n)}\in \H_{R,B}^n$ and $\varphi^{(n)}=0$ for 
$n$ large enough. The vacuum state is $\Omega = (1,0,...,0,...)$. 

At this point we define on $\F_{R,B;m}^0(\H )$ the 
annihilation and creation operators $\{\a , \ac \, :\, f\in \H \}$ 
setting $\a \Omega = 0$ and   
$$
\left [\a \varphi \right ]_{\alpha_1\cdots \alpha_n}^{(n)}(x_1,...,x_n) = 
\sqrt{n+1} \int d^sx\, f^{\dagger \alpha_0 } (x)
\varphi_{\alpha_0 \alpha_1 \cdots \alpha_n}^{(n+1)} (x, x_1,...,x_n) 
\quad , \neweq 
$$
$$
\left [\ac \varphi \right ]_{\alpha_1\cdots \alpha_n}^{(n)}(x_1,...,x_n) 
= \sqrt {n} \left [P^{(n)}_{R,B}\, f\otimes \varphi^{(n-1)}
\right ]_{\alpha_1\cdots \alpha_n}(x_1,...,x_n) \quad , 
\neweq 
$$
{}for all $\varphi \in \F_{R,B;m}^0(\H )$. The operators $\a $ and $\ac $ 
are in general unbounded on $\F_{R,B;m}^0(\H )$. However, for any 
$\psi^{(n)}\in \H_{R,B}^n$ one has the estimates 
$$
\parallel \a \psi^{(n)} \parallel \, \leq \, \sqrt{n} \parallel f 
\parallel \parallel \psi^{(n)} \parallel \, \, , \quad \quad 
\parallel \ac \psi^{(n)} \parallel \, \leq \, \sqrt{n} \parallel 
P^{(n+1)}_{R,B} \parallel \parallel f \parallel \parallel 
\psi^{(n)} \parallel \, \, , \neweq 
$$
$\parallel \, \cdot \, \parallel$ being the $L^2$-norm. 
Therefore $\a$ and $\ac $ are bounded on each $\H_{R,B}^n$. 

The right hand side of eq.(3.27) can be given an alternative form 
by implementing explicitly the action of $P^{(n)}_{R,B}$. The resulting 
expression is a bit complicated, but since in some cases it 
might be instructive, we give it for completeness: 
$$
\left [\ac \varphi \right ]_{\alpha_1\cdots \alpha_n}^{(n)}(x_1,...,x_n) =
{1\over 2 \sqrt{n}} \bigl [f_{\alpha_1}(x_1)\varphi_{\alpha_2\cdots 
\alpha_n}^{(n-1)}(x_2,\dots ,x_n) + 
$$
$$
C(x_1;x_2,...,x_n)_{\alpha_1\cdots \alpha_n}^{\beta_1\cdots \beta_n} 
f_{\beta_1}(-x_1)\varphi_{\beta_2\cdots \beta_n}^{(n-1)}(x_2,\dots ,x_n) 
\bigr ] +  
$$
$$
{1\over 2 \sqrt{n}}\, \sum_{k=2}^{n}  
\left[R_{{k-1}\,k}(x_{k-1},x_k) \cdots R_{1\,2}(x_{1},x_k)  
\right]_{\alpha_1\cdots \alpha_n}^{\beta_1 \cdots 
\beta_n}\bigl [ f_{\beta_1}(x_k)\varphi_{\beta_2\cdots \beta_n}^{(n-1)} 
(x_1,\dots,{\widehat x_k},\dots,x_n) + 
$$
$$  
C(x_k;x_1,...,{\widehat x_k},...,x_n)^{\gamma_1\cdots 
\gamma_n}_{\beta_1\cdots \beta_n} f_{\gamma_1}(-x_k)
\varphi_{\gamma_2\cdots \gamma_n}^{(n-1)} 
(x_1,\dots,{\widehat x_k},\dots,x_n) \bigr ]  
\quad , \neweq 
$$ 
where 
$$
C(x_k;x_1,...,{\widehat x_k},...,x_n)_{\alpha_1\cdots 
\alpha_n}^{\beta_1\cdots \beta_n} = 
$$
$$
\bigl [R_{12}(x_k,x_1)R_{23}(x_k,x_2)\cdots 
{\widehat R_{k\, (k+1)}(x_k,x_k)}\cdots 
R_{(n-1)\, n}(x_k,x_n)B_n(x_k) \cdot 
$$
$$
R_{(n-1)\, n}(x_n,-x_k)\cdots 
{\widehat R_{k\, (k+1)}}(x_k,-x_k)\cdots 
R_{23}(x_2,-x_k) R_{12}(x_1,-x_k)
\bigr ]_{\alpha_1\cdots \alpha_n}^{\beta_1\cdots \beta_n}
\, . \neweq 
$$

We turn now to the boundary generators, defining $b_\alpha^\beta(x)$ 
as the multiplicative operator whose action on $\F_{R,B;m}^0(\H )$ 
is given by eq.(3.7) and 
$$
\left [b_\alpha^\beta(x) \varphi 
\right ]_{\gamma_1 ... \gamma_n}^{(n)}(x_1,...,x_n) = 
[R_{01}(x,x_1) \, R_{12}(x,x_2)\cdots R_{(n-1)\, n}(x,x_n)\, B_n(x)\cdot 
$$
$$ 
\cdot R_{(n-1)\, n}(x_n,-x)\cdots R_{12}(x_2,-x)\, R_{01}(x_1,-x) 
]_{\alpha\gamma_1 ... \gamma_n}^{\beta\delta_1 ... \delta_n} 
\varphi^{(n)}_{\delta_1 ... \delta_n}(x_1,...,x_n) \quad , \neweq 
$$
{}for $n\geq 1$. Notice that the boundary generators 
$\{b_\alpha^\beta(x)\}$ preserve the particle number. 

By construction $\{\a ,\, \ac \}$ and $\{b_\alpha^\beta(x)\}$ 
leave invariant $\F_{R,B;m}^0 (\H )$, which we take as 
the domain $\D$, whose existence was required in the definition 
of Fock representation. For deriving the commutation properties on $\D$ 
it is convenient to introduce the operator-valued distributions 
$a_\alpha (x)$ and $a^{\ast \alpha }(x)$ defined by 
$$
\a = \int d^sx\, f^{\dagger \alpha }(x)a_\alpha (x) \quad , \quad \quad 
\ac = \int d^sx\, f_\alpha (x)a^{\ast \alpha }(x) \quad . \neweq 
$$
After a straightforward but lengthly computation, one verifies the 
validity of the following statement.  
\medskip 
\noindent {\bf Proposition 5.} {\it The operator-valued distributions} 
$\{a_\alpha (x), a^{\ast \alpha }(x) \}$ {\it and} $\{b_\alpha^\beta(x)\}$ 
{\it satisfy the relations} (2.3-8) {\it on} $\D$. 
\medskip 

Assuming that $\M(R,B) \not= \emptyset $, we proceed further by 
implementing the involutions $\{I_m\, :\, m\in \M(R,B) \}$. For 
this purpose we have to construct a sesquilinear form $\form $ 
on $\D$, such that the mapping (2.12-14) is realized 
as the conjugation with respect $\form $. Let us consider 
the following form on $\D$: 
$$
\langle \varphi , \psi \rangle_m\, = \sum_{n=0}^\infty 
\langle \varphi^{(n)} , \psi^{(n)}\rangle_m \quad , \neweq 
$$
where 
$$
\langle \varphi^{(0)} , \psi^{(0)} \rangle_m\, = 
{\overline \varphi^{(0)}}\psi^{(0)} 
\quad , \neweq 
$$
$$
\langle \varphi^{(n)} , \psi^{(n)} \rangle_m\, = 
$$
$$
\int dx_1\cdots dx_n \varphi^{(n)\dagger \alpha_1 ... \alpha_n} 
(x_1,...,x_n)m^{\beta_1}_{\alpha_1}(x_1)\cdots 
m^{\beta_n}_{\alpha_n}(x_n)
\psi^{(n)}_{\beta_1 ... \beta_n}(x_1,...,x_n) \quad . \neweq 
$$
The right hand side of (3.33) always makes sense because for any 
$\varphi , \psi \in \D$ the series is 
actually a finite sum. Using that $m(x)$ satisfies eqs.(2.15) and 
(3.6), one easily proves 
\medskip 
\noindent {\bf Proposition 6.} {\it The inner product defined by} 
(3.33-35) {\it is nondegenerate on} $\D$ {\it and the involution} 
$I_m$ {\it is implemented by} $\form $-{\it conjugation}. 
\medskip 
 
The next question concerns the positivity of $\form $. 
This point is conveniently discussed after introducing the subset 
$\M(R,B)_+$ of those elements of $\M(R,B)$, which are positive 
definite almost everywhere in $\R^s$. One has indeed 
\medskip 
\noindent {\bf Proposition 7.} {\it The inner product} $\form $ 
{\it is positive definite on} $\D$ {\it if and only if} $m\in \M(R,B)_+$. 
\medskip 
\noindent {\it Proof.} From eq.(3.35) it is clear that if  
$m\in \M(R,B)_+$ then the inner product is positive definite. 
Conversely, suppose that $\form $ is positive definite. 
Let $ y\in\R^s $ be a fixed non zero vector, and take an arbitrary 
$f\in \H$ with support laying in the half space $ x\cdot y \geq 0$. 
Consider the 1-particle state 
$$
\varphi_\alpha(x) = [P^{(1)}_{R,B} f]_\alpha (x) = 
{1\over 2} \left [ f_\alpha(x)+B_\alpha^\beta(x) f_\beta(-x) \right ]
\quad . \neweq
$$
Using eqs. (3.6,10) and the support properties of $f_\alpha$, one gets 
$$
\langle \varphi\, , \, \varphi\rangle_m\, = 
{1\over 2} \int d^sx \, f^{\dagger\alpha}(x) m_\alpha^\beta(x) f_\beta(x) 
\quad . \neweq 
$$
Since $f$ is arbitrary, positivity of $\form $ implies 
that $m(x)$ is positive definite almost everywhere in the 
half space $ x\cdot y \geq 0$. Finally, the arbitrariness of $y$ 
allows to extend the validity of this conclusion to $\R^s$. 

Proposition 7 shows that there are two kinds of Fock 
representations of $\balg $. The representation $\F_{R,B;m}$ will be 
called of type A if $\form $ is positive definite; otherwise we will say 
that $\F_{R,B;m}$ is of type B. The standard probabilistic interpretation 
of quantum field theory applies directly only to the A-series. This 
does not mean however that the B-series has no physical applications. 
In the last case one has to isolate first a physical subspace where 
$\form $ is nonnegative. This is usually done by symmetry consideration 
and may depend on the specific model under consideration. 

The final step in completing the derivation of $\F_{R,B;m}$ 
is the construction of the representation space $\L$. 
It is necessary at this stage to consider the classes A and B separately. For 
$m\in \M(R,B)_+$ the inner product space $\{\D, \form \}$ 
is actually a pre-Hilbert space. Let $\F_{R,B;m} (\H )$ 
be the completion of $\D$ with respect 
to the Hilbert space topology. Clearly $\L = \F_{R,B;m} (\H )$ 
satisfies all the requirements.  

{}For type B representations there is no distinguished Hilbert space 
topology for completing $\D$. A natural 
substitute is the topology $\tau $ defined by the family of 
seminorms 
$$
s_\psi (\varphi ) \equiv |\langle \psi \,  ,\, \varphi \rangle_m | \quad , 
\quad \quad \varphi \, , \psi \in \D \quad . \neweq 
$$
It turns out [2] that $\tau $ is the weakest locally convex topology in which 
$\form $ is separately $\tau$-continuous. Moreover, $\tau $ is a Hausdorff 
topology, because $\form $ is nondegenerate. Therefore $\D$ 
admits a unique (up to isomorphism) $\tau$-completion, which has all the 
needed properties and provides the space $\L$ for the B-series. 

We conclude this section by a general observation, which  
concerns A-type representations only and is based on 
the fact that any $m\in \M(R,B)_+$ can be written in the form 
$m(x) = p^\dagger (x)\, p(x)$, where $p(x)$ is an invertible matrix. 
Notice that $p(x)$ is not unitary unless $m(x) = e$. It is easy 
to show that the mapping induced by   
$$
a_\alpha (x) \longmapsto p^\beta_\alpha (x)\, a_\beta (x) \quad , 
\quad \quad 
a^{\ast \alpha (x)} \longmapsto a^{\ast \beta }(x)\, 
p^{-1 \alpha}_{\, \, \, \, \, \, \, \beta}(x)  
\quad , \neweq  
$$
$$
b_\alpha^\beta  (x) \longmapsto 
p_\alpha^\gamma (x)\, b_\gamma^\delta (x)\, 
p^{-1 \beta}_{\, \, \, \, \, \, \, \delta}(-x)  \neweq 
$$
is an isomorphism between $\{\balg ,\, I_m\}$ and 
$\{\B_{R^\prime },\, I_e\}$, where 
$$
R^{\prime \beta_1 \beta_2}_{\, \, \alpha_1 \alpha_2 }(x_1,x_2) = 
p_{\alpha_1}^{\gamma_1}(x_1)\, p_{\alpha_2}^{\gamma_2}(x_2)\, 
R^{\delta_1 \delta_2}_{\gamma_1 \gamma_2 }(x_1,x_2)\, 
p^{-1 \beta_1}_{\, \, \, \, \, \, \, \delta_1}(x_2)\, 
p^{-1 \beta_2}_{\, \, \, \, \, \, \, \delta_2}(x_1) 
\quad . \neweq 
$$
Setting 
$$
B_{\, \, \, \alpha}^{\prime \beta }(x) = p_\alpha^\gamma (x)\, 
B_\gamma^\delta (x)\, p^{-1 \beta}_{\, \, \, \, \, \, \, \, \delta}(-x) 
\quad , \neweq 
$$
one has in addition that $\F_{R,B;m}$ and $\F_{R^\prime ,B^\prime ;e}$ are 
equivalent. In other words, for any $m\in \M(R,B)_+$ one can 
equivalently replace $I_m$ with $I_e$, suitably modifying (see eqs.(3.41,42)) 
the exchange factor $R$ and the reflection matrix $B$. 

Let us mention finally that the above formalism carries over 
easily to the Fock representations of $\balgl $. 
One must only replace the Lebesgue measure $d^s x$ by the\break  
$\lambda$-invariant measure $ |\det \lambda'(x)|^{1/2} d^s x$. 

\newchapt {Applications} 

\leftline {\bf 4.1. Free Boson Field on the Half Line}

\bigskip

In order to give a first idea about the physical content 
of the algebra $\balg $, we focus below on a simple example of 
quantization in $\hl $. More precisely, we construct 
the free boson field $\Phi (t,x)$, satisfying 
$$
\left( \der_t^2 - \der_x^2 + M^2 \right ) \Phi(t,x) = 0 
\quad , \qquad x \in \hl \quad , \neweq
$$
with the boundary condition
$$
\lim_{x \downarrow 0} (\der_x - \eta) \, \Phi(t,x) = 0
\quad , \quad \quad \eta \geq 0 \quad . \neweq
$$
The standard Neumann and Dirichlet boundary conditions are 
recovered from (4.2) by setting $\eta = 0$ or taking the 
limit $\eta \rightarrow \infty $ respectively. 
 
We will show that the quantization of the system (4.1,2) 
can be described in terms of $\B_R$ with $N=1$ 
and $R=1$. The exchange structure of this boundary algebra is trivial, 
which allows to isolate and easily illustrate the physical implications 
of the boundary generator $b(k)$. In this section 
the arguments of the $\B_R$-generators have the meaning of momenta and 
are denoted therefore by $k, p,$ etc. 

Let us introduce the phase factor 
$$
B(k) = {k - i \eta \over k + i \eta}
\quad . \neweq
$$
Then the triplet $\{ R=1, B; m=e \}$ satisfies all requirements of the previous 
section and one can construct the corresponding Fock representation $\F_{1,B;e}$. 
Eq.(3.30) shows that the operator $b(k)$ acts as a multiplication by $B(k)$. 
Therefore, one is left in $\F_{1,B;e}$ with the following relations: 
$$
\eqalign{
[a(k)\, ,\, a(p)] & = 0 \quad, \cr
[a^*(k)\, ,\, a^*(p)] & = 0 \quad, \cr
[a(k)\, ,\, a^*(p)] & = {1\over 2} \delta(k-p)+ {1\over 2} B(k) \delta(k+p)
\quad . \cr} \neweq
$$
Notice that these would be the standard canonical commutation relations, 
apart from the term $B(k)\delta(k+p)$. We define now the field operator
$$
\Phi(t,x)=\int_{-\infty}^\infty {dk\over \sqrt{2\pi \omega(k)}}
\left[ a(k) \, \e^{-i\omega(k) t + i k x} + 
a^*(k) \, \e^{ i\omega(k) t - i k x} \right]
\quad , \neweq 
$$
where 
$$ 
\omega(k)=\sqrt{M^2 + k^2}
\quad . \neweq
$$
This is just the expression in the case without boundary, but 
one should keep in mind that now the algebra of creation and annihilation 
operators is different. 

By means of (4.4) one easily derives the basic 
correlator - the two-point Wightman function 
$$
\langle \Omega \, ,\,  \Phi(t_1,x_1)\Phi(t_2,x_2)\Omega \rangle_e =  
\int_{-\infty }^\infty {dk\over 4\pi \omega(k) }
\e^{-i\omega(k) t_{12}}
\left [  \e^{-i k (x_1-x_2)} + 
B(k) \e^{ - i k (x_1+x_2)} \right ] , \neweq
$$
where $t_{12} = t_1-t_2$. The right hand side of eq.(4.7) 
defines a tempered distribution ($B(k)$ is $C^\infty$ and bounded on $\R$), 
which satisfies eqs.(4.1,2). It consists of two terms. The 
term without $B(k)$ is the usual two-point Wightman function
of the system without boundary. The term proportional to $B(k)$ 
has its origin in the boundary generator and 
explicitly breaks translation and Lorentz invariance. It is remarkable 
that in spite of this fact, $\Phi (t,x)$ is a local field. The validity of 
this statement can be deduced from the commutator 
$$
[\,\Phi(t_1,x_1) \, ,\, \Phi(t_2,x_2)\,]= i D(t_1-t_2,x_1,x_2)
\quad . \neweq
$$
One has 
$$
D(t,x_1,x_2)= \Delta (t,x_1-x_2) + 
{\widetilde \Delta}(t,x_1+x_2)
\quad, \neweq
$$
where 
$$
\Delta (t,x_1-x_2) = 
-\int_{-\infty}^\infty {dk\over 2\pi {\omega(k)}} \,
\sin[\omega (k)t] \, \e^{i k (x_1-x_2)} \neweq
$$
is the ordinary Pauli-Jordan function with mass $M$ and 
$$
\widetilde \Delta (t,x_1+x_2) = 
-\int_{-\infty}^\infty {dk\over 2\pi {\omega(k)}} \,
\sin[\omega (k)t] \, B(k) \, \e^{i k (x_1+x_2)} 
\quad . \neweq 
$$
Observing that for $x_1,\, x_2\in \R_+$ the inequality 
$|t_1-t_2|< |x_1-x_2|$ implies\break $|t_1-t_2|< x_1 + x_2$, 
one concludes that the locality properties of
the field $\Phi$ are governed by the behavior of 
${\widetilde \Delta}(t,x)$ for $|t| < x $. The latter can 
be easily evaluated and using that $\eta \geq 0$, one finds 
$$
{\widetilde\Delta}(t,x)\vert_{{}_{|t| < x }} = 0
\quad . \neweq 
$$
So, $\Phi (t,x)$ is a local field when $x \in \hl $. Notice that this is 
not the case if $\Phi (t,x)$ is considered on the whole real line. 
The two terms $\Delta$ and ${\widetilde\Delta}$ in the commutator 
have a very intuitive explanation. As far as 
$|t_1-t_2|<|x_1-x_2|$ no signal can propagate between 
the points $(t_1,x_1)$ and $(t_2,x_2)$ and the commutator vanishes. When\break 
$|x_1-x_2|<|t_1-t_2|< x_1 + x_2 $ signals can propagate 
directly between the two points, but they cannot be influenced 
by the boundary and the only contribution comes from the standard 
Pauli-Jordan function $\Delta$. As soon as $ x_1 + x_2 = |t_1-t_2|$, 
signals starting from one of the points can be reflected 
at the boundary and reach the other point. This phenomenon is 
responsible for the term ${\widetilde\Delta}$, and is codified in term 
proportional to $B(k)$ of the boundary algebra (4.4). 

The case $\eta < 0$ is slightly more delicate due to the 
presence of a bound state in the one-particle energy spectrum, 
which must be taken into account in the construction of a 
local field. 

The results of this subsection can be obviously generalized to 
higher space-time dimensions.

\vskip 1.5 truecm

\leftline {\bf 4.2. Scattering on the Half Line} 

\bigskip  

Before entering the details of the application of 
$\balg $ to factorized scattering with 
reflecting boundary conditions, we will discuss the simple 
case of particles of mass $M$ freely moving on $\hl $ 
and bouncing over a wall at $x=0$. The relevant 
one-particle space is $L^2(\R_+, dx)$. We denote by 
$D_\eta \subset L^2(\R_+, dx)$ the subspace of 
$C^\infty $-functions on $\R_+$, which vanish for 
sufficiently large $x$, have square integrable first and 
second derivatives and obey 
$$
\lim_{x\downarrow 0}\left ( {d\over dx} - \eta \right ) 
\varphi (x) = 0 \quad . \neweq 
$$
The current 
$$
j = -{i\over 2m}\left [ {\overline \varphi }{d\varphi \over dx} 
- {d\, {\overline \varphi }\over dx}\varphi \right ]  \neweq 
$$
satisfies $j(0) = 0$ for all $\varphi \in D_\eta $, thus 
preventing any probability flow through the wall $x=0$. For 
one-particle Hamiltonian we take 
$$
H^{(1)} = -{1\over 2M} \triangle \quad , \neweq 
$$
defined on $D_\eta $. The evolution problem is well posed 
because $H^{(1)}$, which is obviously symmetric, is actually 
essentially self-adjoint [22]. A set of (generalized) 
eigenstates verifying (4.13) is 
$$
\psi_k(x) = \e^{-ikx} + B(k)\e^{ikx} \quad , 
\quad \quad k \in \R \quad , \neweq 
$$
where $B(k)$ is given by eq.(4.3). The eigenvectors (4.16), 
which represent physically scattering states, satisfy 
$$
\psi_{-k}(x) = {\overline \psi}_k(x) = B(-k)\psi_k(x) 
\quad . \neweq 
$$
{}For $\eta \geq 0$ the systems $\{\psi_k\, :\, k>0\}$ and 
$\{\psi_{-k}\, :\, k>0\}$ are separately complete and are 
related via complex conjugation, which in the physical context 
implements time reversal. When $\eta <0$, there is in 
addition a unique bound state 
$$
\psi_{\rm b}(x) = \sqrt {-2\eta }\, \e^{\eta x} \quad , \neweq 
$$
with energy $E=-\eta^2/2M$. 

The $n$-body Hamiltonian of the associated multiparticle 
Bose system 
$$
H^{(n)} = -{1\over 2M} (\triangle_1 + ... +\triangle_n) \neweq 
$$ 
is defined on $D_{\eta +}^n$ - the subspace of symmetric functions 
in $D_\eta^{\otimes n}$. Clearly, there is neither 
particle production nor particle collision in this model. 
There is however a nontrivial reflection from the boundary, 
which can be described as follows. One can consider  
$\psi_k$ as representing a particle, which when time 
$t\to -\infty$, travels with momentum $-k$ towards the wall. 
Accordingly, we take 
$$
|-k\rangle^{\rm in} = {1\over \sqrt {2\pi}}\, \psi_k (x) \quad , 
\quad \quad  k>0 \quad , \neweq 
$$ 
as a basis of one-particle ``in"-states. Concerning 
the basis of one-particle ``out"-states, the analogous 
consideration gives 
$$
|k\rangle^{\rm out} = 
{1\over \sqrt {2\pi}}\, {\overline \psi}_k (x) = 
{1\over \sqrt {2\pi}}\, \psi_{-k} (x)\quad , 
\quad \quad  k>0 \quad . \neweq 
$$ 
The scattering operator is defined at this point by 
$$
S\, |k\rangle^{\rm out} = |-k\rangle^{\rm in} \quad . \neweq 
$$
{}For $\eta \geq 0$, $S$ is by construction a unitary operator 
on $L^2(\R_+ , dx)$. For $\eta < 0$, $S$ is defined and unitary 
on the subspace of $L^2(\R_+ , dx)$ which is orthogonal to the 
bound state (4.18). The one-particle matrix elements of $S$ read 
$$
{}^{\rm out}\langle k|S|p\rangle^{\rm out} = 
{}^{\rm out}\langle k|-p\rangle^{\rm in} = 
{1\over 2\pi}\int_0^\infty dx \psi_k (x) \psi_p (x) = 
B(k)\delta (k-p) \quad . \neweq 
$$
More generally 
$$
{}^{\rm out}\langle k_1,...,k_n|-p_1,...,-p_n\rangle^{\rm in} = 
B(k_1)...B(k_n)\delta (k_1-p_1)...\delta (k_n-p_n) 
\quad , \neweq 
$$
provided that $k_1>...>k_n>0$ and $p_1>...>p_n>0$. 

Our main observation now is that the above simple scattering problem 
admits a field-theoretic solution in terms of the algebra (4.4). 
In fact, it is easy to verify that the vacuum expectation values 
$$
2^n \langle a^\ast (k_1)...a^\ast (k_n)\Omega \, , 
\, a^\ast (-p_1)...a^\ast (-p_n)\Omega \rangle_e \quad , 
\neweq 
$$
in the Fock representation $\F_{1,B;e}$ reproduce precisely 
the transition amplitudes (4.24). We have therefore 
the following Fock realization 
$$
|k_1,...,k_n\rangle^{\rm out} = 
2^{n\over 2}a^\ast (k_1)...a^\ast (k_n)\Omega 
\quad , \quad \quad k_1>...>k_n>0 \quad , \neweq 
$$
$$
|-p_1,...,-p_n\rangle^{\rm in} = 
2^{n\over 2}a^\ast (-p_1)...a^\ast (-p_n)\Omega \quad , 
\quad \quad p_1>...>p_n>0 \quad , \neweq 
$$
of the interpolating states. Summarizing, the scattering operator 
of our simple model has a purely algebraic characterization. 
In this respect, the term proportional to $B(k)$ in (4.4) 
is the algebraic counterpart of the boundary condition, 
given analytically by eq.(4.13). 

At this stage we have enough background for facing 
the more complicated problem of scattering in 
integrable models with reflecting boundary conditions in 1+1 
space-time dimensions. The presence of particle 
collisions in this case leads in general to the boundary 
algebras $\balg $ with $R\not= 1$. Using the Fock 
representations of $\balg $, derived in the 
previous section, we present below a rigorous construction 
of the $S$-matrix, which generalizes some previous results 
[20] valid in the absence of a boundary. We also show that under 
certain conditions on the triplet $\{R, B;m\}$, 
the transition amplitudes, originally derived by 
Cherednik [4], are indeed Hilbert 
space matrix elements of a unitary operator. 

The asymptotic particles of integrable models 
are parametrized by their rapidity $\theta \in \R $ and 
internal ``isotopic" index $\alpha = 1,...,N$. 
We recall that in the case of relativistic dispersion relation 
the energy-momentum vector is expressed 
in terms of $\theta $ and the mass $M$ according to 
$$
p^0 = M\, {\rm cosh}(\theta )\, \, , \quad \quad p^1 = 
M\, {\rm sinh}(\theta )\, \,  . \neweq 
$$
An elastic reflection $(p^0,\, p^1)\longmapsto (p^0,\, -p^1)$ 
corresponds therefore to the transformation 
$\theta \longmapsto -\theta $. 

The fundamental building blocks for constructing the scattering 
operator are the matrices 
$R_{\alpha_1 \alpha_2 }^{\beta_1 \beta_2 }(\theta_1 , \theta_2 )$ 
and $B_\alpha^\beta (\theta )$, which are supposed to 
satisfy eqs.(2.9,10) and\break (3.9,10). We allow for $R$ to 
depend on $\theta_1$ and $\theta_2$ separately (and not only on 
$\theta_1 - \theta_2$), because in general the 
presence of boundaries brakes down Lorentz invariance. 

A crucial observation is that the algebra $\balg $ alone does not 
determine the scattering operator $S$ we are looking for: one must fix 
in addition an involution $I_m$. The latter selects a Fock 
representation $\brep $, which is the main ingredient for 
constructing $S$. Postponing the discussion of the physical 
meaning of the choice of $m\in \M(R,B)$ to the end of 
this section, it might be instructive for the time being to 
describe the set $\M(R,B)$ for some familiar integrable model. 
We choose the $SU(2)$ Thirring model. In this case $N=2$ and setting 
$\theta_{12}=\theta_1-\theta_2$ the relevant $R$-matrix reads [1] 
$$
R(\theta_1,\theta_2) = 
{i\pi \rho (\theta_{12})\over (i\pi - \theta_{12}) \rho (-\theta_{12})} 
\sum_{\alpha , \beta = 1}^2 
\biggl [ E_{\alpha \alpha }\otimes E_{\beta \beta } + 
{\theta_{12} \over i\pi} (-1)^{\alpha + \beta } 
E_{\alpha \beta }\otimes E_{\beta \alpha }\biggr ]\, \, , \neweq 
$$
where $E_{\alpha \beta}$ are the Weyl matrices and 
$$
\rho (\theta ) = 
\Gamma \left ({1\over 2} + {\theta \over 2\pi i}\right )\,  
\Gamma \left (1- {\theta \over 2\pi i}\right )  \quad . \neweq 
$$
The general solution of eqs.(3.9,10), subject to the physical 
constraint of boundary crossing symmetry [10], is given in [3]. 
Let us concentrate for simplicity on the diagonal solutions 
$$
B(\theta) = {\beta (\theta)\over \beta (-\theta) } 
\left (E_{11} + {\eta - \theta \over \eta + \theta}E_{22}\right ) 
\quad, \neweq
$$
with $\eta \in \C$ and 
$$
\beta (\theta) = \Gamma 
\left ( {3\over 4} + {\theta \over 2\pi i} \right )
\Gamma 
\left ( 1 - {\theta \over 2\pi i} \right )
\Gamma 
\left ( { {\eta + i \pi } - \theta \over 2\pi i} \right )
\Gamma 
\left ( { {\eta + 2 \pi i } + \theta\over 2\pi i} \right )
\quad. \neweq
$$
Let $\mu_+$ ($\mu_-$) be any measurable real-valued even 
(odd) function, such that $\mu_\pm $ and\break 
$1/\mu_\pm $ are bounded. Then, if ${\rm Re}\, \eta = 0$, 
the set $\M(R,B)$ contains all matrices of the form
$$
m(\theta) = \mu_+ (\theta)\left (E_{11} + \xi E_{22}\right ) 
\quad , \quad \quad \xi \in \R ,\, \, \xi \not=0 \quad . \neweq 
$$
In addition, for $\eta = 0$ one has the solutions 
$$
m(\theta) = \mu_- (\theta) 
\left (\zeta E_{12} + \bar \zeta E_{21}\right ) 
\quad, \quad \quad \zeta \in \C \quad . \neweq
$$
{}From eq.(4.33) it follows that $\M(R,B)_+ \not= \emptyset $.

After this concrete example illustrating the set $\M(R,B)$, we 
return to the general framework. The idea is to extend the formalism, 
developed at the beginning of this section for the 
Schr\"odinger particle on the half line, to the case of 
integrable  models. In what follows we assume that 
$$
\M(R,B)_+ \not= \emptyset \neweq 
$$
and consider representations $\brep $ of type A. The physical 
motivation for this restriction is quite evident. According to 
proposition 7, it ensures positivity of the metric in the 
asymptotic spaces $\F^{{\rm out}}$ and $\F^{{\rm in}}$, which we 
are going to construct now. For this purpose we introduce 
the following relation in $C_0^\infty (\R )$: 
$$ f_1 \succ f_2 \quad \Longleftrightarrow \quad  \theta_1 > \theta_2 
\quad \forall \, \theta_1 \in {\rm supp}(f_1)\, , \quad 
\forall \, \theta_2 \in {\rm supp}(f_2) \quad . \neweq 
$$
We will adopt also the notation 
$$ f \succ 0 \quad \Longleftrightarrow \quad  \theta > 0 
\quad \forall \, \theta \in {\rm supp}(f) \quad , \neweq 
$$
and
$$
\widetilde f(\theta ) = f(-\theta ) \quad . \neweq 
$$
As suggested by eqs.(4.26,27), $\F^{{\rm out}}$ and $\F^{{\rm in}}$ 
are generated by finite linear combinations of the vectors ($k\geq 1$)  
$$
\E^{{\rm out}} = 
\{\, \Omega ,\,  a^\ast (f_1) \cdots a^\ast (f_k) \, \Omega \, \, : 
\, \, f_{1_{\alpha_1 }}\succ \cdots \succ f_{k_{\alpha_k }}\succ 0,\, 
\forall \, \alpha_1 ,...,\alpha_k = 1,...,N \, \} \neweq 
$$
and 
$$
\E^{{\rm in}} = 
\{\, \Omega ,\, a^\ast (\widetilde g_1) \cdots 
a^\ast (\widetilde g_k) \, \Omega \, \, : \, \, 
g_{1_{\beta_1 }}\succ \cdots \succ g_{k_{\beta_k }}\succ 0,\, 
\forall \, \beta_1 ,...,\beta_k = 1,...,N \, \} \neweq 
$$
respectively. By construction both $\F^{{\rm out}}$ 
and $\F^{{\rm in}}$ are linear subspaces 
of the Hilbert space $\F_{R,B;m}(\H )$. 

One should notice that in principle there are elements 
of $\F_{R,B;m}^0(\H )$ which belong neither to $\F^{{\rm out}}$ 
nor to $\F^{{\rm in}}$. We call them mixed vectors. Linear 
combinations involving both in- and out-states provide in general 
examples of such vectors. In spite of the existence of mixed vectors, 
the subspaces $\F^{{\rm out}}$ and $\F^{{\rm in}}$ 
satisfy a sort of asymptotic completeness, which is essential for 
constructing the $S$-matrix. More precisely, one has 
\medskip 
\noindent {\bf Proposition 8}. $\F^{{\rm out}}$ {\it and} $\F^{{\rm in}}$ 
{\it separately are dense in} $\F_{R,B;m}(\H )$. 
\medskip 
\noindent {\it Proof}: We focus on $\F^{{\rm out}}$. 
Let $\varphi \in \F_{R,B;m} (\H )$ and let us assume that 
$$
\langle \varphi  \, ,\, \psi \rangle_m = 
0 \quad \forall \, \psi \in \F^{{\rm out}}
\quad . \neweq 
$$
In order to prove the thesis, we have to show that 
$\varphi = \left ( \varphi^{(0)}, \varphi^{(1)},...,\varphi^{(n)},...\right ) 
= 0$. 
Obviously $\varphi^{(0)}=0$. Let us consider $\varphi^{(n)}$ for 
arbitrary but fixed $n\geq 1$. Eq.(3.27) and eq.(4.41) imply that   
$$
\langle \varphi^{(n)}\, ,
\, a^\ast (f_1) \cdots a^\ast (f_n) \Omega \rangle_m = 
$$
$$
\int d\theta_1 \cdots d\theta_n\, \varphi^{(n)\dagger \alpha_1 ... \alpha_n} 
(\theta_1,...,\theta_n) m^{\beta_1 }_{\alpha_1}(\theta_1 ) \cdots 
m^{\beta_n}_{\alpha_n}(\theta_n ) f_{1_{\beta_1 }}(\theta_1 ) 
\cdots f_{n_{\beta_n }} (\theta_n ) = 0 \neweq 
$$
{}for all $f_1,...,f_n$ such that 
$f_{1_{\alpha_1 }}\succ \cdots \succ f_{n_{\alpha_n }}\succ 0\, \, 
 \forall \, \alpha_1 ,...,\alpha_n = 1,...,N$. Therefore 
$$
\varphi^{(n)}_{\alpha_1 ... \alpha_n} (\theta_1,...,\theta_n) = 0 
\neweq 
$$
in the domain $\theta_1>\cdots >\theta_n > 0$. Finally, using that 
$\varphi^{(n)} \in \H_{R,B}^n$ has definite exchange and reflection 
properties described by eqs.(3.24,25), one can extend the domain of 
validity of (4.43) and conclude that $\varphi^{(n)}$ actually vanishes 
almost everywhere in $\R^n$. Clearly, a similar argument applies 
also to the case of $\F^{{\rm in}}$. 
 
We observe in passing that the definition of $\F^{{\rm out}}$ 
and $\F^{{\rm in}}$ does not explicitly involve the boundary generators 
$\{b_\alpha^\beta (\theta )\}$. This fact is not surprising because 
$\Omega $ is cyclic with respect to $\{a^{\ast \alpha }(\theta )\}$. 

At this point we are ready to define the scattering matrix $S$ 
and to prove that it is a unitary operator in $\F_{R,B;m}(\H )$. 
The construction consists essentially of three steps. One starts 
by defining $S$ as the following mapping of $\E^{{\rm out }}$ onto 
$\E^{{\rm in }}$: 
$$
S\Omega = \Omega \quad , \neweq 
$$
$$
S\, a^\ast (g_1) a^\ast (g_2)\cdots a^\ast (g_k)\Omega
= a^\ast (\widetilde g_1) a^\ast (\widetilde g_2)\cdots 
a^\ast (\widetilde g_k)\Omega 
\quad , \neweq 
$$
where $g_{1_{\beta_1 }}\succ \cdots \succ g_{k_{\beta_k }}\succ 0,\, \, 
\forall \, \beta_1 ,...,\beta_k = 1,...,N $. It is not 
difficult to check that 
$$
\langle S\psi^{\rm out}\, ,\, S\varphi^{\rm out}\rangle_m \, 
=\, \langle \psi^{\rm out}\, ,\, \varphi^{\rm out}\rangle_m  \quad , 
\quad \quad \forall \, \psi^{\rm out} ,\, \varphi^{\rm out} \in \E^{\rm out} 
\quad . \neweq 
$$
Moreover, $S$ is invertible and 
$$
\langle S^{-1}\psi^{\rm in}\, ,\, S^{-1}\varphi^{\rm in}\rangle_m \, 
=\, \langle \psi^{\rm in}\, ,\, \varphi^{\rm in}\rangle_m  \quad , 
\quad \quad \forall \, \psi^{\rm in} ,\, \varphi^{\rm in} \in \E^{\rm in} 
\quad . \neweq 
$$

The second step is to extend $S$ and $S^{-1}$ by linearity to 
the whole  $\F^{{\rm out}}$ and $\F^{{\rm in}}$ respectively. 
Clearly, one has to show that these extensions are correctly defined. 
Consider for instance $S$ and suppose that there exist a sequence 
$$
g^i_{1_{\beta_1 }}\succ \cdots \succ g^i_{k_{\beta_k }}\succ 0\, , \quad  
\forall \, \beta_1 ,...,\beta_k = 1,...,N\, , \, \, \, \, \, i = 1,...,M\, , 
$$ 
such that 
$$
a^\ast (g_1) a^\ast (g_2)\cdots a^\ast (g_k)\Omega = \sum_{i=1}^M 
a^\ast (g^i_1) a^\ast (g^i_2)\cdots a^\ast (g^i_k)\Omega  
\quad . \neweq 
$$
In order to prove that the linear extension of $S$ is not ambiguous, 
we must show that 
$$
a^\ast (\widetilde g_1) a^\ast (\widetilde g_2)\cdots 
a^\ast (\widetilde g_k)\Omega = \sum_{i=1}^M 
a^\ast (\widetilde g^{\, i}_1) a^\ast (\widetilde g^{\, i}_2)\cdots 
a^\ast (\widetilde g^{\, i}_k)\Omega  \quad . \neweq 
$$
The argument is as follows. In the domain 
$\theta_1 > \theta_2 > ... > \theta_k > 0$  eq.(4.48) implies that 
$$
g_{1_{\beta_1 }}(\theta_1 )\, g_{2_{\beta_2 }}(\theta_2 ) 
\cdots g_{k_{\beta_k }}(\theta_k ) = 
\sum_{i=1}^M g^i_{1_{\beta_1 }}(\theta_1 )\, g^i_{2_{\beta_2 }}(\theta_2 ) 
\cdots g^i_{k_{\beta_k }}(\theta_k ) 
\quad . \neweq 
$$
Because of the support properties of $\{g_j\}$ and $\{g_j^i\}$ one has that 
eq.(4.50) holds actually in $\R^k$, which projected by $P_{R,B}^{(k)}$ 
proves the validity of eq.(4.49). 

It is easy to see also that eqs.(4.46,47) remain valid for the 
linear extensions of $S$ and $S^{-1}$ on $\F^{{\rm out}}$ and 
$\F^{{\rm in}}$ respectively. This fact implies in particular 
that both $S$ and $S^{-1}$ are bounded linear operators. 

{}Finally, one extends $S$ and $S^{-1}$ by continuity to 
$\F_{R,B;m} (\H )$. Because of the asymptotic completeness 
proven in proposition 8, the 
extensions are unique and define the unitary  scattering 
operator and its inverse. As it should be expected from 
integrability, one has $S\H^n_{R,B} \subset \H^n_{R,B}$.
Notice however, that in contrast to 
the case without boundary, where the scattering operator 
leaves invariant each one-particle state, the $S$-matrix 
constructed above acts nontrivialy already in $\H^1_{R,B}$. 

By construction the matrix elements of $S$ between out-states 
in the Fock space $\F_{R,B;e}(\H)$ reproduce precisely the transition 
amplitudes derived by Cherednik [4]. Since the latter are referred 
to the involution $I_e$, a natural question 
arising at this point concerns the physical meaning of other 
possible choices of $m\in \M(R,B)_+$. For answering this question 
we consider two generic asymptotic 
states $\varphi^{\rm in}\in \F^{\rm in}$ and 
$\psi^{\rm out}\in \F^{\rm out}$. 
If both $m,\, e\in \M(R,B)_+$, one may compare the 
transition amplitudes associated with the involutions $I_m$ and 
$I_e$. One finds 
$$
\langle \psi^{\rm out}\, ,\, \varphi^{\rm in}\rangle_m \, =\,  
\langle \psi^{\rm out}\, ,\, \varphi_d^{\rm in}\rangle_e \, = \, 
\langle \psi_d^{\rm out}\, ,\, \varphi^{\rm in}\rangle_e  
\quad , \neweq 
$$
where $\varphi_d^{\rm in} $ and 
$\psi_d^{\rm out}$ 
are the ``dressed" in- and out-states 
$$
(\varphi^{\rm in}_d)^{(n)}_{\alpha_1...\alpha_n}(\theta_1,...,\theta_n)
= m^{\gamma_1}_{\alpha_1 } (\theta_1 )\cdots 
m^{\gamma_n}_{\alpha_n }(\theta_n )  
(\varphi^{\rm in})^{(n)}_{\gamma_1...\gamma_n}(\theta_1,...,\theta_n) 
\quad , \neweq 
$$ 
$$
(\psi^{\rm out}_d)^{(n)}_{\beta_1...\beta_n}
(\theta_1,...,\theta_n) = 
m^{\dagger \gamma_1}_{\, \, \, \beta_1 }(\theta_1 )\cdots 
m^{\dagger \gamma_n}_{\, \, \, \beta_n }(\theta_n )  
(\psi^{\rm out})^{(n)}_{\gamma_1...\gamma_n}(\theta_1,...,\theta_n) 
\quad . \neweq 
$$ 
It follows from eq.(4.51) that the effect of the involution $I_m$ 
is exactly reproduced in $\F_{R,B;e}$ by appropriate 
dressing (4.52,53) of the in- or out-states. 

The results of this section can be summarized as follows. 
\medskip 
\noindent {\bf Proposition 9.} {\it Suppose that the exchange factor} 
$R$ {\it and the reflection matrix} $B$ {\it satisfy} (2.9,10) 
{\it and} (3.9,10). {\it Assume also that} $\M(R,B)_+\not= \emptyset$. 
{\it Then the scattering operator associated with 
the Fock representation} $\F_{R,B;m}$ {\it is unitary for any} 
$m\in \M(R,B)_+$. 
\medskip

Conditions (2.9,10) and (3.9,10) are standard for the scattering on the 
half line. The same is true for (2.15), which is usually imposed in 
the slightly stronger form 
$$
R^{\dagger \beta_1 \beta_2 }_{\, \, \alpha_1 \alpha_2} (\theta_1,\theta_2) 
= R^{\beta_1 \beta_2 }_{\alpha_1 \alpha_2} (\theta_2, \theta_1) 
\quad , \neweq 
$$
known as Hermitian analyticity. We emphasize that condition (3.6), 
which is often overlooked in the physical literature, 
is essential for the unitarity of $S$ and represents 
therefore an useful criterion for selecting possible reflection 
matrices. In the case of the $SU(2)$ Thirring model one 
gets in this way the restriction ${\rm Re}\, \eta = 0$ in eqs.(4.31,32). 

Let us mention also that if $R$ depends on the difference 
$\theta_{12} \equiv \theta_1 -\theta_2$, one usually assumes [13,26]
that $R$ admits a suitable continuation to the complex $\theta_{12}$-plane, 
which satisfies crossing symmetry, has certain pole structure, etc. 
In that case also $B$ is required to have a continuation in the 
complex $\theta $-plane, which obeys boundary crossing 
symmetry [10]. In our example (see eqs.(4.29-32)) $R$ and $B$ admit 
such continuations. Finally, the bootstrap equations [9,26] 
reduce further the set of physically relevant exchange and 
reflection matrices. From proposition 9 it follows however that 
the unitarity of $S$ as an operator in $\F_{R,B;m}(\H )$ 
depends exclusively on the behavior of $R$ and $B$ for real 
values of the rapidities.

\newchapt {Outlook and Conclusions}

In the present paper we have introduced the associative 
algebra $\balg $ and investigated some of its basic features. 
$\balg $ admits two series of Fock representations, which 
have been constructed explicitly. The positive metric 
representations provide a framework for deriving Cherednik's 
transition amplitudes and proving that they are indeed 
the matrix elements of a uniptary scattering operator. 
We have shown also that the algebra $\B_+$ enters the Bose 
quantization on the half line. The associated Klein-Gordon field 
is local, in spite the breakdown of the Poincar\'e symmetry. 

$\balg $ is actually a member of a large family of 
algebras $\balgl $, which are defined by eqs.(2.23-26). 
$\balgl $ can be studied in the same way as $\balg $ and are expected 
to find relevant applications to statistical models with 
boundaries. It will be interesting in this respect to extend 
to $\balgl $ the notion of second $R$-quantization, developed 
in [17,18] for the Z-F algebra $\alg $.  

We point out finally that one can further generalize $\balgl $, 
eliminating the condition 
(2.9) and/or (3.10). In this case, 
instead with the Weyl group ${\cal W}_n$, one has to 
deal with an infinite dimensional group ${\cal W}_n^\prime $, 
which is freely generated by the elements 
$\{\tau^\prime ,\, \sigma_i^\prime \, \, :\, \, i=1,..., n-1\}$ 
satisfying the relations (3.14,15), but not (3.16). 
Recent investigations [15] show actually that the group 
${\cal W}_n^\prime $ appears in many different physical and 
mathematical contexts. We hope to 
say more about this generalization of $\balgl $ in the near future. 

\vskip 1.5 truecm

\centerline {\bbf Appendix}

\bigskip

In quantum field theory on the half line it 
is sometimes necessary to allow for a quantum number 
$j = 1,...,N_B$ to reside on the boundary [10]. 
We will show below that this case is still described by the 
boundary algebra $\{\balg , I_m\}$, but corresponds to 
representations with slightly more general structure then that of $\brep $. 
To be precise, instead of the requirement 4 formulated in the beginning 
of Sect. 3, these representations satisfy: 

\medskip
\item {$4^\prime $.} There exists a $N_B$-dimensional subspace (vacuum space) 
$\V \subset \D$, which is annihilated by $a_\alpha (x)$. 
Moreover, $\V$ is cyclic with respect to $\{a^{\ast \alpha } (x)\}$ and 
$\form $ is positive definite on $\V$. 
\medskip

\noindent For $N_B = 1$ we recover the property 4 specifying $\brep $. 

Let us briefly describe now the main features of the 
representations characterized by the conditions 1-3 and $4^\prime $. 
Let $\Omega^1,\ldots , \Omega^{N_B}$ be an orthonormal basis 
in $\V$. We denote by $P_0$ be the $\form $-orthogonal projection on $\V$ 
and define
$$
B_\alpha^\beta (x) \equiv P_0 \, b_\alpha^\beta (x) \, P_0 
\quad . \eqno (A.1) 
$$
Notice that $B_\alpha^\beta (x)$ is now an operator, 
carrying the vacuum space into itself, 
$$
B_\alpha^\beta (x) \, \Omega^j = 
B_{\alpha \, k}^{\beta \, j}(x) \, \Omega^k \quad . \eqno (A.2) 
$$
The following obvious generalization of Proposition 3 holds. 

\medskip
\noindent {\bf Proposition $3^\prime $.} {\it The vacuum space} $\V $ 
{\it is unique} {\it and satisfies} 
$$
b_\alpha^\beta (x)\, \vert_{\V} =  B_\alpha^\beta (x)\, \vert_{\V} 
\quad . \eqno (A.3)  
$$
\medskip

\noindent Projecting the relevant equations on the vacuum space, 
one immediately verifies the validity of (3.6,9,10) as operator equations on $\V$. 

Summarizing, the basic input for constructing the above more general class 
of representations 
of $\{\balg , I_m\}$ is still the triplet $\{R,B;m\}$, the novelty being 
that $B_\alpha^\beta(x)$ are operators which satisfy (3.6,9,10) 
on $\V$. Apart from the following minor modifications, the construction 
precisely follows that described in Sect. 3. First of all, the elements of 
$\H^n_{R,B}$ carry an extra lower index varying from 1 to $N_B$. In 
the scalar product this index is saturated among the two states. Second, 
performing the substitution 
$B_{\alpha_i }^{\beta_i }(x) \mapsto B_{\alpha_i \, k}^{\beta_i \, j}(x)$ 
in eq.(3.20), the operator $B_i(x)$ becomes a $N_B \times N_B$-matrix, 
which inserted in (3.30,31) acts on the states by a standard matrix 
multiplication.

\vfill\eject

\centerline {{\bbf References}} 

\medskip 

\item {1.} Belavin, A. A.: Exact Solution of the Two-dimensional 
Model with Asymptotic Freedom. Phys. Lett. B {\bf 87} 117-123 (1979) 

\item {2.} Bogn\'ar, J.: Indefinite Inner Product Spaces.  Berlin: 
Springer Verlag, 1974 

\item {3.} Chao, L.,  Hou, B.,  Shi, K., Wang, Y., Yang, W.: 
Bosonic Realization of Boundary Operators in $SU(2)$-invariant 
Thirring Model. Int. J. Mod. Phys. A {\bf 10}, 4469-4482 (1995) 

\item {4.} Cherednik, I. V.: Factorizing Particles on a Half Line 
and Root Systems. Theor. Math. Phys. {\bf 61}, 977-983 (1984) 

\item {5.} Corrigan, E., Dorey, P. E., Rietdijk, H. R.: Aspects 
of Affine Toda Field Theory on a Half Line. Suppl. Prog. Theor. Phys. 
{\bf 118}, 143-164 (1995)

\item {6.} Faddeev, L. D.: Quantum Completely Integrable Models 
in Field Theory. Soviet Scientific Reviews Sect. C {\bf 1}, 
107-155 (1980)

\item {7.} Fendley, P., Saleur, H.: Deriving Boundary S Matrices. 
Nucl. Phys. B {\bf 428}, 681-693 (1994) 

\item {8.} Fring, A., K\"oberle, R.: Affine Toda Field Theory in 
the Presence of Reflecting Boundaries. Nucl. Phys. B {\bf 419}, 647-664 
(1994) 

\item {9.} Fring, A., K\"oberle, R.: Factorized Scattering in the 
Presence of Reflecting Boundaries. Nucl. Phys. B {\bf 421}, 159-172 
(1994) 

\item {10.} Ghoshal, S., Zamolodchikov, A. B.: 
Boundary S Matrix and Boundary State in Two-Dimensional 
Integrable Quantum Field Theory. Int. J. Mod. Phys. 
{\bf A9}, 3841-3886 (1994)  

\item {11.} Ghoshal, S.: Boundary S-matrix of the $O(N)$-Symmetric 
Non-Linear Sigma Model. Phys. Lett. B {\bf 334}, 363-368 (1994) 
 
\item {12.} Ghoshal, S.: Bound State Boundary S-Matrix of the 
Sine-Gordon model. Int. J. Mod. Phys. A {\bf 9}, 4801-4810 (1994) 

\item {13.} Karowski, M., Weisz, P.: Exact Form Factors in 
(1+1)-Dimensional Field Theoretic Models with Soliton Behavior. 
Nucl. Phys. B {\bf 139}, 455-476 (1978) 

\item {14.} Kuczma, M., Choczewski, B., Ger, R.: Iterative Functional 
Equations. Cambridge: Cambridge University Press, 1990 

\item {15.} Kulish, P. P., Sasaki, R.: Covariance Properties 
of Reflection Equation Algebras. Progr. Theor. Phys. {\bf 89}, 
741-761 (1993) 

\item {16.} LeClair, A., Mussardo, G., Saleur, H., Skorik, S.: 
Boundary Energy and Boundary States in Integrable Quantum Field 
Theories. Nucl. Phys. B {\bf 453}, 581-618 (1995) 

\item {17.} Liguori, A., Mintchev, M.: Fock Representations of 
Quantum Fields with Genera\-lized Statistics. Commun. Math. Phys. 
{\bf 169}, 635-652 (1995) 

\item {18.} Liguori, A., Mintchev, M., Rossi, M.: Unitary Group 
Representations in Fock Spaces with Generalized Exchange 
Properties. Lett. Math. Phys. {\bf 35}, 163-177 (1995)

\item {19.} Liguori, A., Mintchev, M.: Boundary Exchange Algebras. 
Preprint IFUP-TH 21/96, March 1996

\item {20.} Liguori, A., Mintchev, M., Rossi, M.: Fock 
Representations of Exchange Algebras with Involution. 
J. Math. Phys. {\bf 38} 2888-2898 (1997)

\item {21.} Penati, S., Zanon, D.: Quantum Integrability in 
Two-Dimensional Systems with Boundary. Phys. Lett. B {\bf 358}, 
63-72 (1995) 

\item {22.} Reed, M.,Simon, B.: Methods in Modern Mathematical 
Physics II: Fourier Analysis, Self-Adjointness. New York: 
Academic Press, 1975

\item {23.} Saleur, H., Skorik, S., Warner, N.P.: 
The Boundary Sine-Gordon Theory: Classical and Semi-Classical 
Analysis. Nucl. Phys. B {\bf 441}, 421-436 (1995)

\item {24.} Sklyanin, E. K.: Boundary Conditions for Integrable Quantum 
Systems. J. Phys. A: Math. Gen. {\bf 21}. 2375-2389 (1988)

\item {25.} Warner, N. P.: Supersymmetry in Boundary Integrable Models. 
Nucl. Phys. B {\bf 450}, 663-694 (1995)

\item {26.} Zamolodchikov, A. B., Zamolodchikov, A. B.: Factorized 
$S$-Matrices in Two Dimensions as the Exact Solutions of Certain 
Relativistic Quantum Field Theory Models. Ann. Phys. {\bf 120}, 
253-291 (1979)

\item {27.} Zhao, L.: Fock Spaces with Reflection Condition and 
Generalized Statistics. Pre\-print hep-th 96040024

\vfill\eject 
\bye